\documentclass[reprint,amsmath,nofootinbib,amssymb,aps,prc]{revtex4-2}
\usepackage{graphicx}
\usepackage{multirow}
\usepackage{epstopdf}
\usepackage{color}
\usepackage{bm}
\usepackage{textcomp} 
\usepackage{subfigure}
\usepackage{dcolumn} 
\usepackage{bm}
\usepackage{float}

\begin{document}

\title{Search for the chiral magnetic wave \\ using anisotropic flow of identified particles at RHIC}
\collaboration{The STAR Collaboration}
\date{\today}

\begin{abstract}

The chiral magnetic wave (CMW) has been theorized to propagate in the deconfined nuclear medium formed in high-energy heavy-ion collisions, and to cause a difference in elliptic flow ($v_{2}$) between negatively and positively charged hadrons. Experimental data consistent with the CMW have been reported by the STAR Collaboration at the Relativistic Heavy Ion Collider (RHIC), based on the charge asymmetry dependence of the pion $v_{2}$ from Au+Au collisions at $\sqrt{s_{\rm NN}}$ = 27 to 200 GeV.
In this  comprehensive study, we present the STAR measurements of elliptic flow and triangular flow of charged pions, along with the $v_{2}$ of charged kaons and protons, as a function of charge asymmetry in Au+Au collisions at $\sqrt{s_{\rm NN}}$ = 27, 39, 62.4 and 200 GeV.
The slope parameters extracted from the linear dependence of the $v_2$ difference on charge asymmetry for different particle species are reported and compared in different centrality intervals. In addition, the slopes of $v_{2}$ for charged pions in small systems, \textit{i.e.}, $p$+Au and $d$+Au at $\sqrt{s_{\rm NN}}$ = 200 GeV, are also presented and compared with those in large systems, \textit{i.e.}, Au+Au at $\sqrt{s_{\rm NN}}$ = 200 GeV and U+U at 193 GeV. Our results provide new insights for the possible existence of the CMW, and further constrain the background contributions in heavy-ion collisions at RHIC energies.
\end{abstract}

\maketitle


\section{Introduction}
\label{sec:intro}
The violation of parity symmetry ($\cal P$) or  combined charge conjugation and parity symmetry ($\cal CP$) in the strong interaction is allowed by quantum chromodynamics, but has never been observed in experiments (see Ref.~\cite{Baker_2006} for the latest
experimental limits). Metastable $\cal P$- and $\cal CP$-odd  domains may exist in the hot and dense nuclear medium created in high-energy heavy-ion collisions, owing to vacuum transitions induced by topologically nontrivial gluon fields, {\it e.g.}, sphalerons~\cite{Kharzeev_1998}. In such domains, a nonzero chirality chemical potential ($\mu_5$) can arise from the chiral anomaly switching the chirality of quarks, {\it e.g.}, left-handed quarks may become right-handed in the presence of the negative topological charge. The chemical potential $\mu_5$, if coupled with an intense magnetic field ($\overrightarrow{B}$), will induce an electric current along $\overrightarrow{B}$ via the so-called chiral magnetic effect (CME)~\cite{Kharzeev_2008,Fukushima_2008, Kharzeev_2016}: $\overrightarrow{J_e} \propto \mu_5\overrightarrow{B}$. The required magnetic field, as strong as $B \sim 10^{15}$~T in Au+Au collisions at the top RHIC energy, can be produced by the energetic spectator protons in noncentral collisions.

A complementary phenomenon to the CME is the chiral separation effect (CSE)~\cite{Son_2004,Metlitski_2005}, whereby a chirality current is induced along $\overrightarrow{B}$ in the presence of a finite electric chemical potential ($\mu_{\rm e}$): $\overrightarrow{J_5} \propto \mu_{\rm e}\overrightarrow{B}$. The CME and the CSE intertwine to form a collective excitation, the chiral magnetic wave (CMW), a long-wavelength hydrodynamic mode of chiral charge densities~\cite{Burnier_2011,Kharzeev_2011_1,Yee_2014,Burnier_2016,Newman_2006}. The CMW is assumed to be a signature of chiral symmetry restoration~\cite{Schafer_1998}, and manifests itself in a finite electric quadrupole moment of the collision system, where the ``poles" and the ``equator" of the produced fireball  acquire additional positive and negative charges, respectively~\cite{Burnier_2011}. This effect, if present, will be reflected in the measurements of a charge-dependent elliptic flow. 

The anisotropic flow quantifies the collective motion of the expanding medium, and is defined in terms of the Fourier coefficients of the azimuthal distribution of produced particles with respect to the $n$th-order event plane, $\Psi_n$~\cite{Poskanzer_1998}:
\begin{equation}
\frac{dN}{d \varphi} \propto 
1 + \sum_{n = 1} ^\infty 2 v_n \cos  n(\varphi-\Psi_n), 
\end{equation}
where $\varphi-\Psi_n$ is the particle's azimuthal angle with respect to the event plane angle. The quantity $v_1$ is known as ``directed flow", $v_2$ as ``elliptic flow" and $v_3$ as ``triangular flow".
The electric quadrupole moment induced by the CMW will lead to the increase (decrease) of $v_{2}$ for negatively (positively) charged hadrons. The modification of $v_{2}$ due to this effect is predicted to be proportional to the event-by-event charge asymmetry ($A_{\rm ch}$)~\cite{Burnier_2011}, a proxy for $\mu_e$,
\begin{equation}
v_{2}^{\pm} - v_{\rm 2, base}^{\pm} = \mp \frac{a}{2}A_{\rm ch},
\end{equation}
where superscript $\pm$ denotes the positively or negatively charged particles, $v_{2,\rm base}$ represents the ``usual" $v_{2}$ unrelated to the charge separation, $a$ is the quadrupole moment normalized by the net charge density, and
\begin{equation}
A_{\rm ch} = (N^{+} -N^{-}) / (N^{+} +N^{-}),  
\end{equation} with {$N^{+}$} ({$N^{-}$}) denoting the number of positive (negative) particles observed in a given event.

Experimental measurements of such a linear dependence between $v_2^{\pm}$ and $A_{\rm ch}$ is quantified by the slope parameter, $r_2 = d\Delta v_{2} / d A_{\rm ch}$, where $\Delta v_{2} = v_{2}^- - v_{2}^+$.
Although recent STAR measurements~\cite{STAR-isobar} observe no predefined CME signatures in the isobar data ($^{96}$Ru+$^{96}$Ru and $^{96}$Zr+$^{96}$Zr), we cannot exclude the possibility that the CMW observable has a better signal-to-background ratio than the CME ones.
Past measurements have been performed with charged pions in Au+Au collisions by the STAR collaboration at RHIC~\cite{Adamczyk_2015} as well as with charged hadrons in Pb+Pb collisions by the ALICE collaboration at the Large Hadron Collider (LHC)~\cite{Adam_2016, Voloshin_2014}. In both cases, the $r_2$ slopes are of the same order of magnitude as predicted by theoretical calculations of the CMW~\cite{Burnier_2011,Kharzeev_2011_1,Yee_2014,Burnier_2016,Newman_2006}. In particular, the STAR results exhibit the expected centrality dependence. However, non-CMW mechanisms could also contribute to the splitting of $v_{2}^\pm$ as a function of $A_{\rm ch}$.
A hydrodynamic study~\cite{Hatta_2016} claims that the simple viscous transport of charges, combined with certain initial conditions, will lead to a sizeable $v_2$ splitting for charged pions. According to the analytical calculation of the anisotropic Gubser flow~\cite{Gubser}, the $\Delta v_{2}$ for pions is proportional to both the shear viscosity and the isospin chemical potential ($\mu_{\rm I}$)~\cite{Hatta_2016, Hatta2015}.
On the other hand, charge asymmetry $A_{\rm ch}$ can also be linearly related to $\mu_{\rm I}$ with the help of a statistical model, which consequently connects $\Delta v_2$ and $A_{\rm ch}$. This model further predicts negative $r_2$ slopes for charged kaons and protons with larger magnitudes than the pion slopes, because  $\mu_{\rm I}$ as well as the strangeness chemical potential $\mu_{\rm S}$ will affect these particles differently. These predictions warrant the extension of our measurements to kaons and protons.  

Local charge conservation (LCC)~\cite{Bzdak_2013, Voloshin_2014, Wu2021, Wang2021} is also able to qualitatively explain the finite $r_2$ slope observed from data, when convoluted with the characteristic dependence of $v_2$ on particle pseudorapidity ($\eta$) and transverse momentum ($p_{T}$). This is demonstrated with locally charge-conserved clusters, {\it e.g.}, a pair of particles with opposite charges, originating from a fluid element or a resonance decay. Such a pair could contribute to a non-zero $A_{\rm ch}$ in an experiment, when one of the particles escapes the limited detector acceptance. If this process preferentially occurs in a phase space with smaller $v_2$, such as a lower-$p_{T}$ or higher-$\eta$ region, then there would be a positive $r_2$ slope, whether the escaping particle is positive or negative. For example, the escape of a $\pi^+$ with smaller $v_2$ effectively increases the $v_2$ of detected $\pi^+$'s, and decreases the observed $A_{\rm ch}$, causing a negative slope for detected $\pi^+$'s. Conversely, the escape of a $\pi^-$ with smaller $v_2$ increases the $v_2$ of detected $\pi^-$'s, and also increases the observed $A_{\rm ch}$, causing a positive slope for detected $\pi^-$'s.
A realistic estimate of such contributions, however, appears to be smaller than that observed in the STAR measurements~\cite{Adamczyk_2015}. Ref.~\cite{Bzdak_2013} also proposes a test with the $r_3$ measurements, defined as $r_3 = d\Delta v_3 / d A_{\rm ch}$ with $\Delta v_3 = v_3^- - v_3^+$, which should yield finite slopes according to the LCC picture, while no slope is expected from the CMW picture. Recently the CMS collaboration at the LHC~\cite{PhysRevC.100.064908} has observed that normalized $r_2$ and $r_3$ slopes are very similar to each other for charged hadrons in Pb+Pb collisions at 5.02 TeV, supporting the LCC picture. Such a test with the STAR data at 200 GeV is reported in this paper.

The CMS measurements~\cite{PhysRevC.100.064908} also show, for charged hadrons, a very similar $A_{\rm ch}$ dependence of $\Delta v_{2}$ in $p$+Pb and Pb+Pb collisions at 5.02 TeV. In $p$+Pb collisions, the magnetic field direction is presumably decoupled from the event plane~\cite{Belmont_2017}, and the $r_2$ slopes are dominated by non-CMW contributions.
The similar $r_2$ slopes in $p$+Pb and Pb+Pb collisions~\cite{PhysRevC.100.064908} suggest that the $r_2$ slopes measured in Pb+Pb  are unlikely to originate from the CMW. This disappearance of the CMW could arise from the fact that the magnetic field strength drops in the vacuum much faster at the LHC energies than at RHIC~\cite{Bfield}, and at the time of quark production, the magnetic field could become too weak to initiate the CMW. The potential difference in the physics mechanisms between RHIC and the LHC motivates us to present STAR measurements of $r_2$ in small systems, \textit{i.e.}, $p$+Au and $d$+Au at 200 GeV, and to compare them with results for Au+Au and U+U collisions.

This paper is organized in the following way.
The STAR experiment and data collection are briefly introduced in Sec.~\ref{sec:star}. The analysis methods and systematic uncertainties are described in Sec.~\ref{sec:anaMth}. The STAR results of the $A_{\rm ch}$ dependence of identified particle anisotropic flow are presented and discussed in Sec.~\ref{sec:results}, where we report: (A) the $A_{\rm ch}$ dependence of mean $p_{T}$  and mean $|\eta|$, and the $\Delta v_{2}$ slope for charged pions selected using different phase space requirements; (B) the $r_2$ slopes for charged kaons and protons; (C) the $r_{3}$ slope for charged pions; (D) the $r_{2}$ slopes for charged pions in $p$+Au, $d$+Au and U+U. A summary is given in Sec.~\ref{sec:sum}.


\section{Experimental setup and data selection} \label{sec:star}

\begin{figure}
\includegraphics[width=0.48\textwidth]{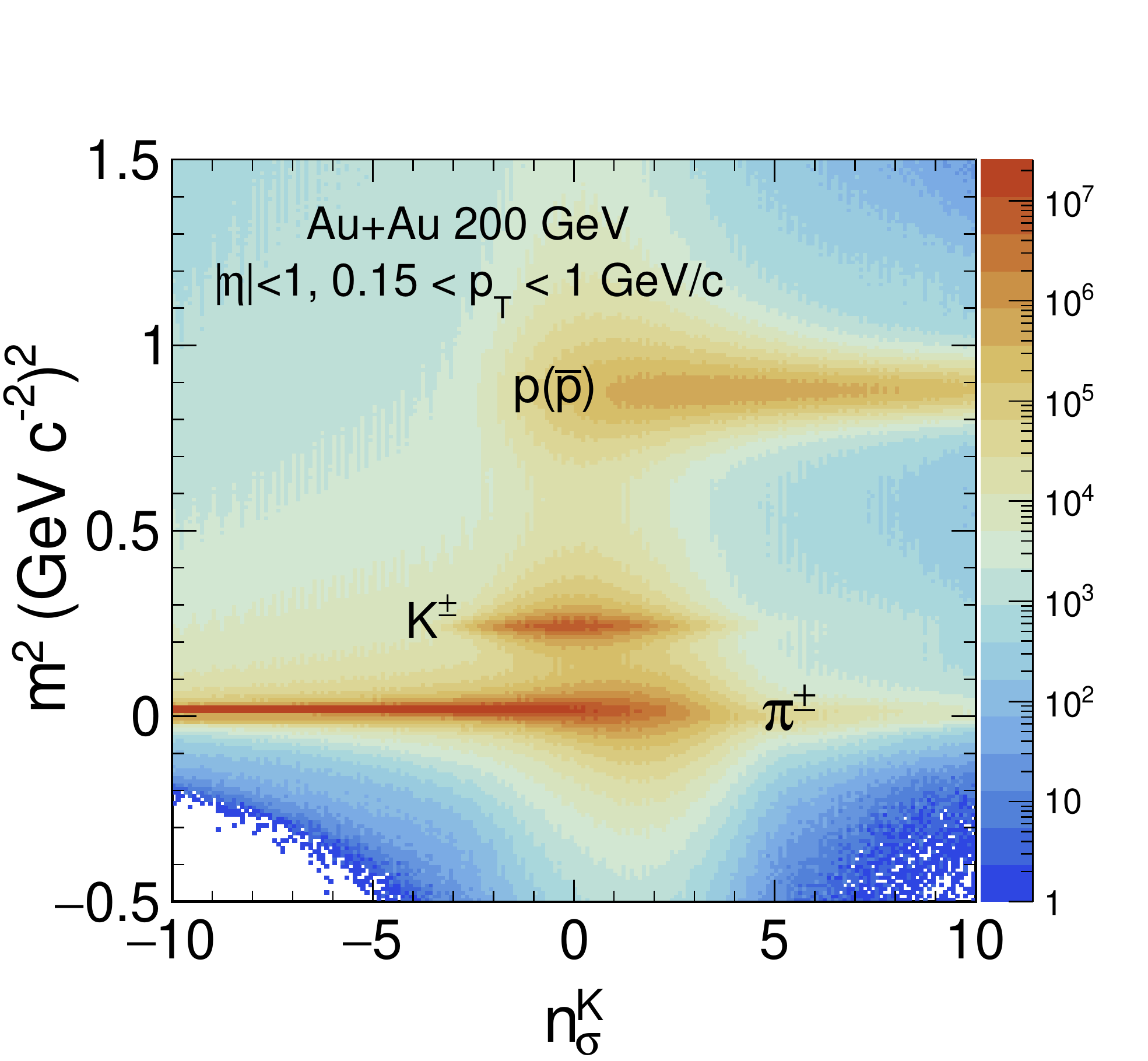}
\caption{Particle identification by the STAR TPC and TOF detectors. 
$n_{\sigma}$ denotes the deviations from the theoretical $ln(dE/dx)$ curves measured by the TPC (here for kaons), while $m^2$ denotes the mass information deduced from  the TOF.}
\label{fig:pid}
\end{figure}

The STAR detector complex consists of a series of subsystems located in both midrapidity and forward-rapidity regions. The main detectors involved in this work are the Time Projection Chamber (TPC)~\cite{Ackermann_2003, Anderson_2003} and the Time-of-Flight (TOF) detector~\cite{Bonner_2003, Llope_2012}. The TPC is surrounded by a solenoidal magnet providing a uniform magnetic field along the beam direction, and tracks charged particles in the pseudorapidity window $|\eta|<1.3$, with full azimuthal angle coverage. The track curvature determines the transverse momentum and the charge sign of the corresponding particle, and its mean ionization energy loss per unit track length ($dE/dx$) is used to identify the particle species. The TOF encloses the curved surface of the cylindrical TPC, and together with momentum from the TPC provides information on the mass of the particle. In this work, particles are jointly identified by the TPC and the TOF, as shown in Fig.~\ref{fig:pid}. The TPC selects $\pi^\pm$, $K^\pm$, $p$ and $\bar{p}$ within a $2\sigma$ window centered on the expected $ln(dE/dx)$ curve for each species. The calculated mass requirements with TOF are $-0.05 < m^{2}_{\pi} < 0.1 ~{\rm GeV}^2/c^4$, $0.15 < m^{2}_{K} < 0.35~{\rm GeV}^2/c^4$, and $0.6 < m^{2}_{p} < 1.2~{\rm GeV}^2/c^4$. 
As a systematic check, the analyses have been repeated using only the TPC $dE/dx$ for particle identification.

The data samples of heavy-ion collisions consist of minimum-bias triggered events taken by the STAR detector, including Au+Au collisions at $\sqrt{s_{\rm NN}}$ = 27, 39, 62.4 and 200 GeV, as well as U+U collisions at $\sqrt{s_{\rm NN}}$ = 193 GeV.
The 27 GeV Au+Au data were collected in the year 2011, the 39 and 62.4 GeV data in 2010, the 200 GeV data in  2011, 2014 and 2016, while the U+U data were collected in the year 2012.
Each data set is divided into nine centrality classes according to the so-called ``reference multiplicity" at STAR, which is generally determined by the raw multiplicity of primary charged particles reconstructed in the TPC over the full azimuth and $|\eta| < 0.5 $. The choice of $|\eta| < 0.5 $ is made because in this region the $\eta$ distribution is almost flat,  and the tracks have better quality than those near the TPC edge.
The
centrality classes are defined by fitting the reference multiplicity distribution to that obtained from MC Glauber
simulations~\cite{glauber1,glauber2}. In Glauber simulations, 
the number of participant nucleons ($N_{\rm part}$)
is obtained by MC sampling.
The centrality definition procedure also determines a multiplicity-dependent weight that is applied to each event to correct for the event reconstruction inefficiency, especially in peripheral collisions. Data of $p$+Au and $d$+Au collisions at 200 GeV come from the years 2015 and 2016, respectively. The primary vertex of each event is required to be within 30 cm from the detector center along the beam direction, and within a radius of 2 cm from the beam in the transverse direction to eliminate background events which involve interactions with the beam pipe. Since the small systems have narrow reference multiplicity distributions, we do not divide those data samples into finer centrality intervals.

A set of track quality cuts was implemented. Tracks are required to have $\geq15$ space points in the TPC fiducial acceptance ($|\eta| < 1$, chosen to match that of the TOF detector), and have a ratio of the number of measured space points over the maximum possible number of space points larger than 0.52, which effectively prevents double-counting of a particle due to track splitting. To prevent inclusion of secondary particles, a track is rejected if its distance of closest approach (DCA) to the primary vertex is larger than 1 cm. The DCA cut was varied in the study of the systematic uncertainty. A minimum transverse momentum of 0.15 GeV/$c$ is also required, because the reconstruction process for tracks with lower $p_{T}$ is hindered by the low tracking efficiency and the limited acceptance. All these selection criteria, both event-wise and track-wise, are consistent with those used in the previous STAR publication on the same topic~\cite{Adamczyk_2015}.


\section{Analysis method} \label{sec:anaMth}

In this analysis, we investigate the $A_{\rm ch}$ dependence of anisotropic flow for various particle species. The measured charge asymmetry, \textit{i.e.}, the ratio of net charge over the total charge multiplicity, is calculated for charged particles with $0.15 < p_{T} < 12$ GeV/$c$, excluding the low-$p_{T}$ protons and antiprotons ($p_{T} < 0.4$ GeV/$c$) to avoid  potential ``knock-out protons" from the beam pipe. For a given centrality, the event sample is divided into five $A_{\rm ch}$ sub-groups, each with similar numbers of events. Owing to the limited detector efficiency, the measured $A_{\rm ch}$ needs to be scaled to match the distribution of the true $A_{\rm ch}$. According to our previous study~\cite{Adamczyk_2015}, the relationship between the measured $A_{\rm ch}$ and the true $A_{\rm ch}$ appears to be almost linear. We calculate and compare the $A_{\rm ch}$ before and after correcting for tracking efficiency with the HIJING~\cite{Wang_1994} and AMPT~\cite{Lin_2002} models. 
For convenience, the tracking efficiency of pions is used for all charged particles, since pions are the dominant particles. The assumed efficiency is varied for systematic checks.  

The Q-cumulant method \cite{Bilandzic_2011} is adopted to extract the anisotropic flow, which provides a fast and accurate calculation without looping over all particle combinations. In this approach, all multi-particle cumulants are expressed with respect to flow vectors (${\bf Q}_n\equiv\sum_{k=1}^Me^{in\varphi_k}$). For instance, two-particle correlations for a single event and for all events, respectively, can be calculated by
\begin{equation}
\langle 2^{'} \rangle = \frac{{\bf p}_{n} {\bf Q}^{*}_{n}-m_{q}}{m_{p}M-m_{q}},
\end{equation}
and
\begin{equation}
\langle\langle 2^{'} \rangle\rangle = \frac{\sum^{N}_{i=1}(w_{\langle 2^{'} \rangle})_{i}\langle 2^{'} \rangle_{i}}{\sum^{N}_{i=1}(w_{\langle 2^{'} \rangle})_{i}},
\label{eq:all_cumu}
\end{equation}
where {${\bf p}_{n}$} and {${\bf Q}_{n}^{*}$} are flow vectors, and $w_{\langle 2^{'} \rangle}$ represents the event weight, \textit{i.e.}, multiplicity. The {$m_{p}$} and $M$ are the number of particles of interest (POI) and the number of reference particles (RFP), respectively, while {$m_{q}$} denotes the number of particles labeled by both POI and RFP. One first estimates
the reference flow by using only the RFPs, and then
obtains the differential flow of POIs with
respect to the reference flow of the RFPs. Using the differential second-order cumulant $d_{n}\{2\}=\langle\langle 2^{'} \rangle\rangle$, one can estimate differential flow by
\begin{equation}
v_{n} \{2\} = \frac{d_{n}\{2\}}{\sqrt{c_{n}\{2\}}},
\label{eq:diff_cumu}
\end{equation}
where $c_{n}$ represents the reference flow calculated in the similar way \cite{Bilandzic_2011}. An {$\eta$} gap of 0.3 is applied between POIs and RFPs to suppress short-range nonflow effects \cite{Adamczyk_2013_2}.

There are two schemes to calculate $\Delta v_n$. (a) Find the $p_{T}$-integrated $v_{n}$ in a given $p_{T}$ range for negatively and positively charged particles, and then take the difference. (b) Start with the $v_{n}$ difference between negatively and positively charged particles as a function of $p_{T}$, and then fit the difference in the specified $p_{T}$ range with a constant to extract the average. We have confirmed that results from these two schemes are consistent with each other, and we choose the former result as the central value and the latter as a systematic check. 

\begin{figure}[h]
\includegraphics[width=20pc]{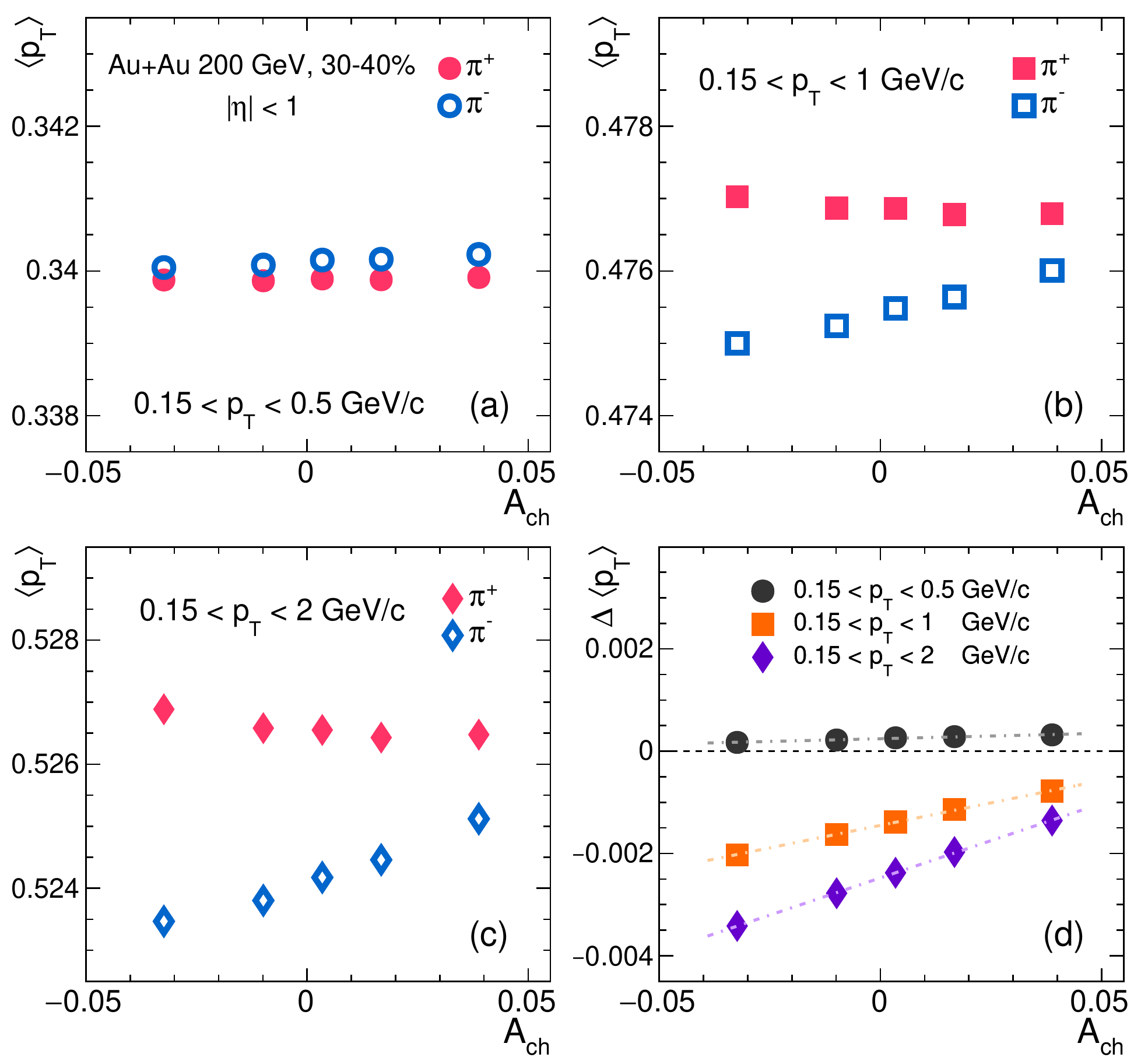}
\caption{$A_{\rm ch}$ dependence of $\langle p_{T} \rangle$ and $\Delta \langle p_{T} \rangle$   for $\pi^{\pm}$ in various $p_{T}$ ranges with $|\eta|<1$ in 30--40\% centrality Au+Au collisions at $\sqrt{s_{\rm NN}}$ = 200 GeV. Uncertainties are only statistical, and are smaller than the marker size. The dashed lines represent the linear fits.}
\label{fig:linear_meanPt}
\end{figure}

\begin{figure}[h]
\includegraphics[width=20pc]{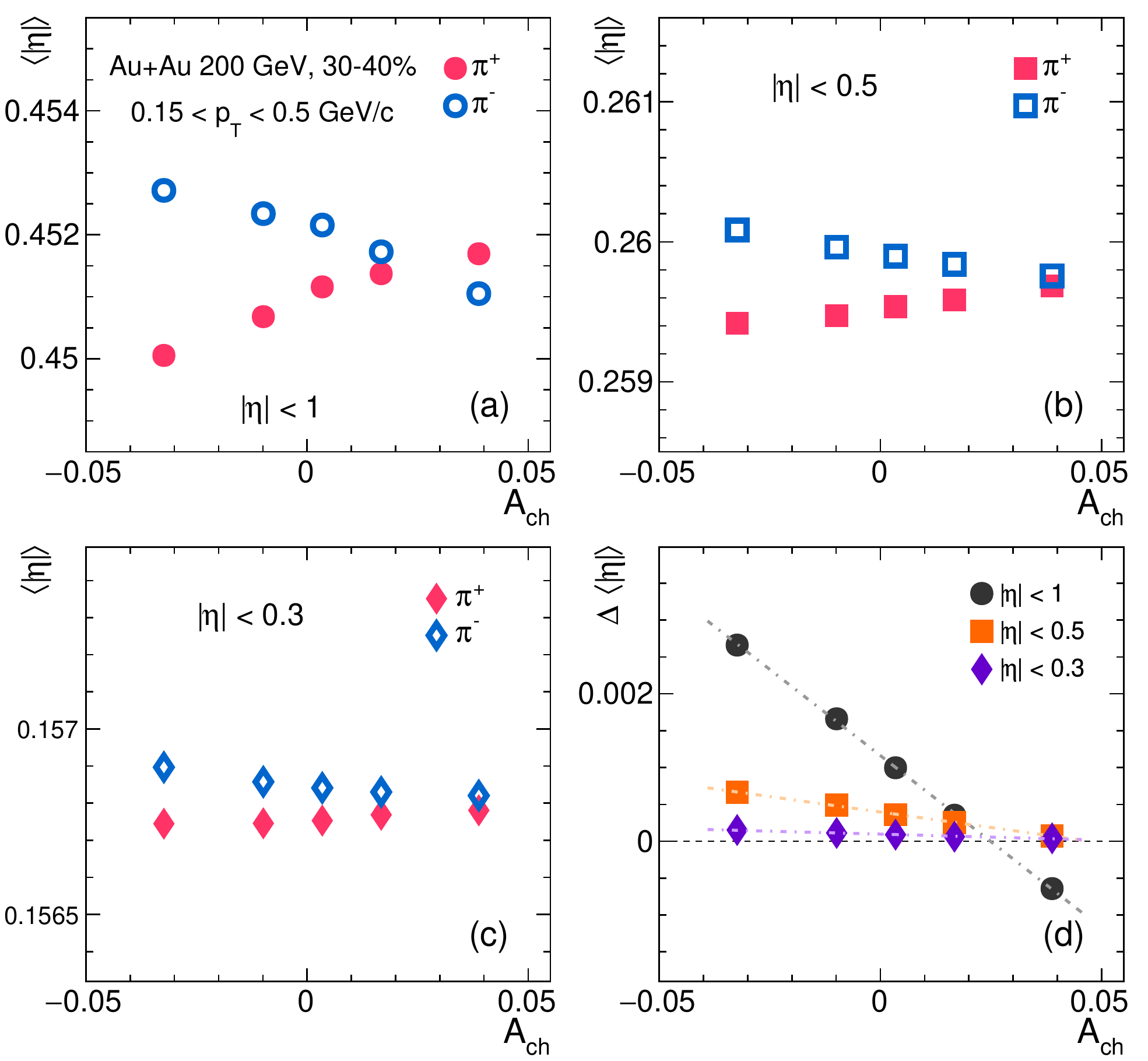}
\caption{$A_{\rm ch}$ dependence of $\langle |\eta| \rangle$ and $\Delta \langle |\eta| \rangle$  for $\pi^{\pm}$ in various $\eta$ ranges with $0.15 < p_T < 0.5$ GeV/$c$ in 30--40\% centrality Au+Au collisions at $\sqrt{s_{\rm NN}}$ = 200 GeV. Uncertainties are only statistical, and are smaller than the marker size. The dashed lines represent the linear fits.}
\label{fig:linear_meanEta}
\end{figure}

\begin{figure}
\includegraphics[width=20pc]{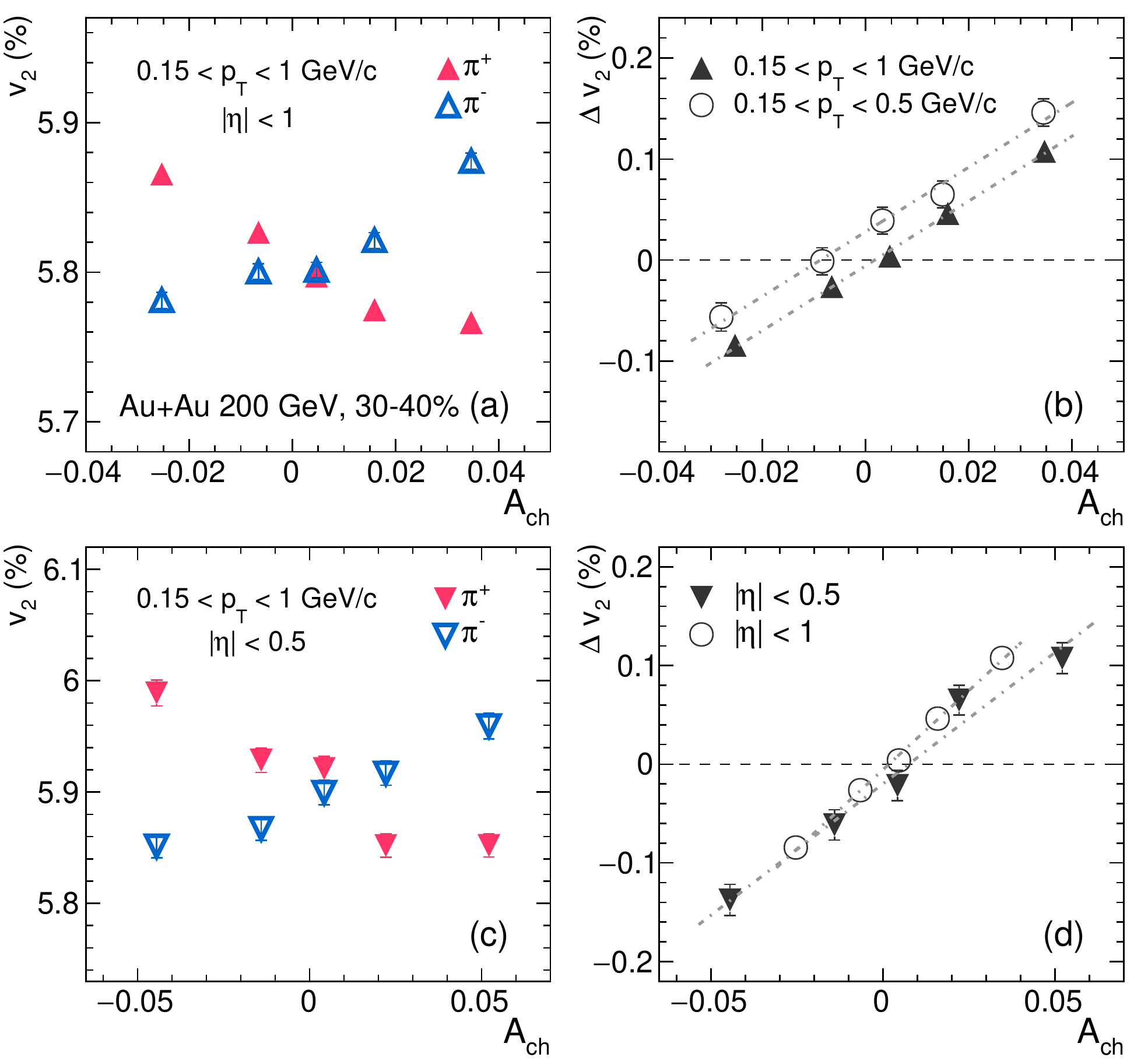}
\caption{$v_2$ for $\pi^\pm$ and $\Delta v_{2}$ in different $p_{T}$ and $\eta$ windows as a function of $A_{\rm ch}$ in 30--40\% centrality Au+Au collisions at $\sqrt{s_{\rm NN}}$ = 200 GeV. Uncertainties are only statistical. The dashed lines represent the linear fits to the given data points.}
\label{fig: linear_pt_eta}
\end{figure}

The systematic uncertainties for each data sample are estimated by varying cuts and extracting $\Delta v_{n}$ with different methods. Here we briefly summarize the systematic sources and their typical contributions.
The uncertainties from particle identification mainly include two sources: altering the DCA cut of POIs from 1 cm to 0.5 cm, as well as identifying the particle with both TPC and TOF or with TPC only. For the case of pions, varying the DCA cut and performing PID with only TPC give rise to $\sim18\%$ and $\sim10\%$ uncertainties, respectively. For the cases of kaons and protons, such cut variations introduce an extra effect of relative 5--10\% owing to the detector performance. The estimation of $A_{\rm ch}$ partly depends on the tracking efficiency of the TPC. Therefore, we accordingly adjust the efficiency by $\pm5\%$ and observe a consequent $\sim10\%$ variation of the slope values. When calculating $\Delta v_{n}$, the discrepancy between the aforementioned two methods
is found to be smaller than $5\%$. In addition, the data sets from different years could also result in different but consistent slope values, which are reflected in the uncertainties too. None of the aforementioned uncertainties shows any significant change when the measurements are performed at different $p_{T}$ ranges.


\section{Results and discussions} \label{sec:results}


\begin{figure}
\includegraphics[width=20pc]{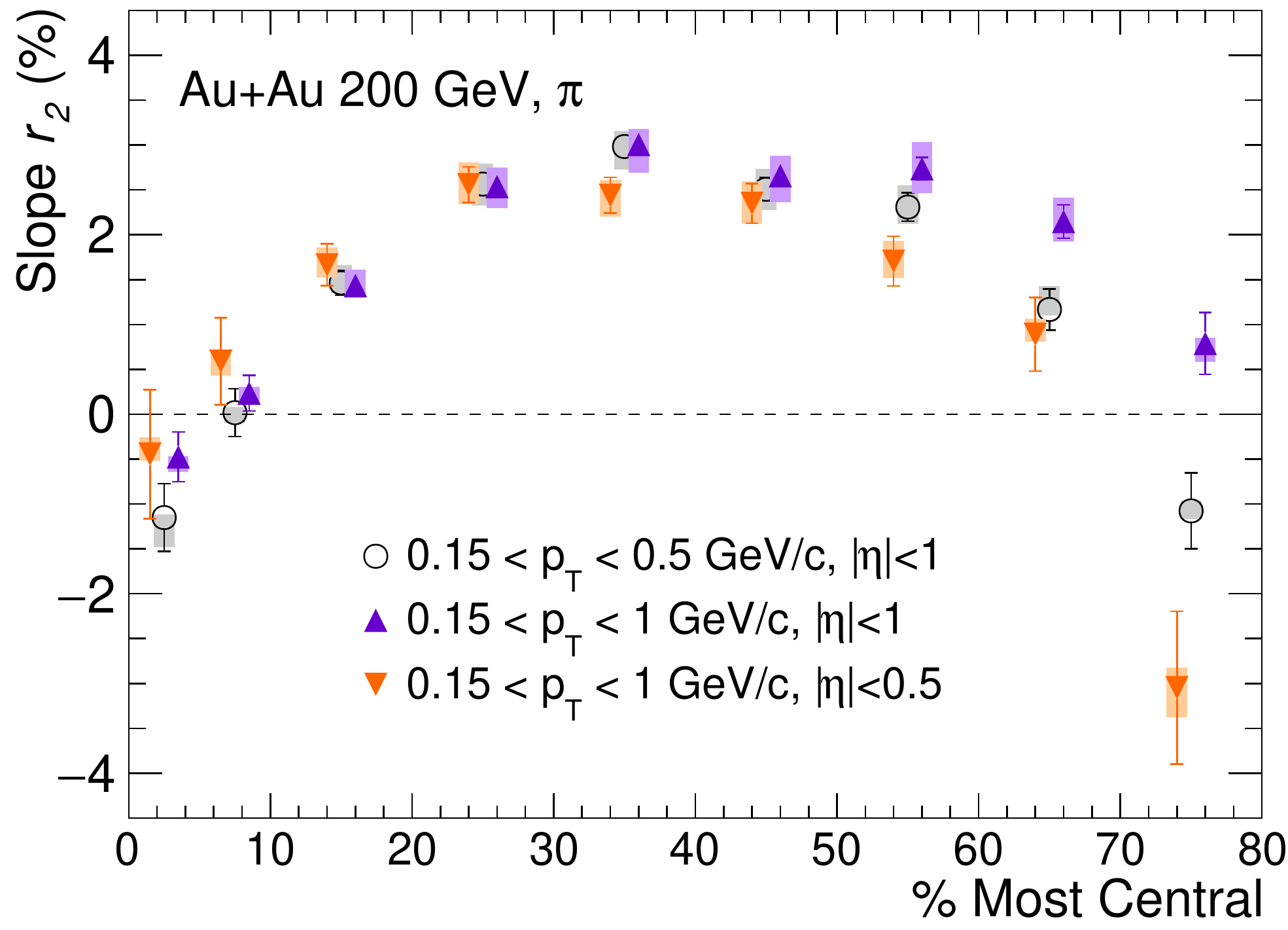}
\caption{
The $r_2$ slopes obtained with different
$p_{T}$ and $\eta$ ranges as a function of centrality in Au+Au collisions at $\sqrt{s_{\rm NN}}$ = 200 GeV. Some points are horizontally shifted for clarity. 
}
\label{fig: slo_pt_eta}
\end{figure}

\begin{figure*}
\includegraphics[width=33pc]{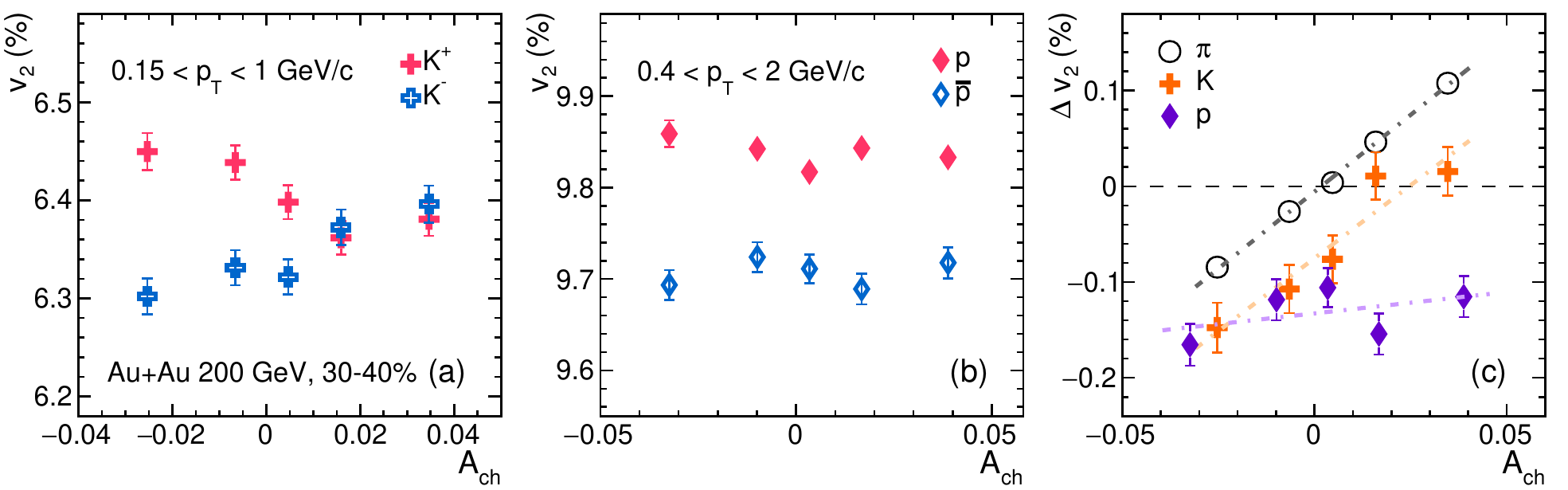}
\caption{$v_2$ for $K^\pm$ (a),  $p$ and $\bar p$ (b) and $\Delta v_{2}$ for pions, kaons and protons (c) as a function of $A_{\rm ch}$ in 30--40\% centrality Au+Au collisions at $\sqrt{s_{\rm NN}}$ = 200 GeV. The dashed lines represent the linear fits to the given data points.}
\label{fig:linear_k_p}
\end{figure*}

\subsection{Dependence of $\langle p_{T} \rangle$ and $v_{2}$ on $A_{\rm ch}$ for pions in different kinematic windows}

The previous STAR measurement~\cite{Adamczyk_2015} examined the dependence of $\Delta v_{2}$ on $A_{\rm ch}$ for pions with $0.15 < p_{T} < 0.5$ GeV/$c$. There are two reasons for this choice of the $p_{T}$ range. First, the CMW is a collective phenomenon, affecting primarily the bulk particles at low momenta. Second, $v_2$ has a strong dependence on $p_{T}$, and if the mean $p_{T}$ of particles, $\langle p_{T} \rangle$, changes with $A_{\rm ch}$, then $v_2$ (and further $\Delta v_{2}$) appears to depend on $A_{\rm ch}$. 
Our goal is to properly select the $p_{T}$ range to reveal the pertinent physics.
Figure~\ref{fig:linear_meanPt} presents $\langle p_{T} \rangle$ for $\pi^{\pm}$ with different $p_{T}$ ranges in panels (a)-(c) and $\Delta \langle p_{T} \rangle$, defined as $\langle p_{T} \rangle^{\pi^-}-\langle p_{T} \rangle^{\pi^+}$, in panel (d) as a function of $A_{\rm ch}$ in the 30--40\% centrality bin for Au+Au collisions at 200 GeV.
The narrow range of $0.15 < p_{T} < 0.5$ GeV/$c$ leads to roughly constant $\langle p_{T} \rangle$ for $\pi^{+}$ and $\pi^{-}$ separately, and $\Delta \langle p_{T} \rangle$ is close to zero regardless of $A_{\rm ch}$.
The other two wider $p_{T}$ ranges yield a stronger dependence of $\langle p_{T} \rangle$ and $\Delta \langle p_{T} \rangle$ on $A_{\rm ch}$. 

As introduced in Sec.~\ref{sec:intro}, in a finite $\eta$ acceptance, the dependence of $v_2$ on $\eta$ could couple with LCC to result in a finite $r_2$ slope~\cite{Bzdak_2013}. 
Recent studies~\cite{Wu2021,Wang2021} again emphasize that the $A_{\rm ch}$ dependence of mean $p_{T}$ and that of mean $|\eta|$ can be directly explained by the LCC.
Figure~\ref{fig:linear_meanEta} shows $\langle|\eta|\rangle$ for $\pi^{\pm}$ with different $\eta$ ranges in panels (a)-(c) and $\Delta \langle |\eta| \rangle$, defined as $\langle |\eta| \rangle^{\pi^-}-\langle |\eta| \rangle^{\pi^+}$, in panel (d) as a function of $A_{\rm ch}$ in the 30--40\% centrality interval.
The default range of $|\eta|<1$ displays the strongest $\langle|\eta|\rangle$ variation ($\sim$0.5\%). 

To study the impact of the $\langle p_{T} \rangle$ variation on the final observables, we show $v_{2}$ for $\pi^{\pm}$ and $\Delta v_{2}$ in different $p_{T}$ windows as a function of $A_{\rm ch}$ in panels (a) and (b) of Fig.~\ref{fig: linear_pt_eta}. The $r_2$ slopes are $3.20(\pm 0.29)\%$ and $3.21(\pm 0.17)\%$ for 0.15 $< p_{T} <$ 0.5 GeV/c and 0.15 $< p_{T} <$ 1 GeV/c, respectively. The increased upper bound of $p_{T}$ has a marginal effect on the $r_2$, because over the same $A_{\rm ch}$ range, the relative variation of $\langle p_{T} \rangle$ ($\sim$0.1\%) is typically smaller than the relative variation of $v_{2}$ ($\sim$1\%) by an order of magnitude, and $v_{2}$ is  roughly proportional to $\langle p_{T} \rangle$.
This result confirms the model study in Ref.~\cite{Wang2021}. It does not mean that the LCC effect has been eliminated because even when the same integral $p_T$ and $\eta$ cuts are applied, the differential kinematic windows could still be different for $\pi^{+}$ and $\pi^{-}$ in the same $A_{\rm ch}$ bin.
On the other hand, a wider $p_{T}$ range enhances particle yields, which is important for analyses involving $K^\pm$,  $p$ and $\bar{p}$. Therefore, an optimal $p_{T}$ range in experiment is needed to take both statistics and systematics into account.
The effect of the $\langle |\eta|\rangle$ variation is investigated via $v_2$ and $\Delta v_2$ for $\pi^\pm$ in different $\eta$ windows, as shown in panels (c) and (d) of Fig.~\ref{fig: linear_pt_eta}. When the $\eta$ coverage is reduced to half, the slope $r_2$ does not display a significant variation.
For the remainder of this study, we simply focus on $|\eta|<1$ unless otherwise stated.

The slope parameters obtained with different phase space selections are compared in Fig.~\ref{fig: slo_pt_eta} as a function of centrality. Note that the result for $p_{T} <$ 0.5 GeV/c is slightly different from the published one~\cite{Adamczyk_2015}, simply owing to the different data sets. 
The results show very similar rise-and-fall trends, except for very peripheral collisions, where  nonflow effects could make a difference to the measurements with different kinematic cuts.

\begin{figure}
\includegraphics[width=20pc]{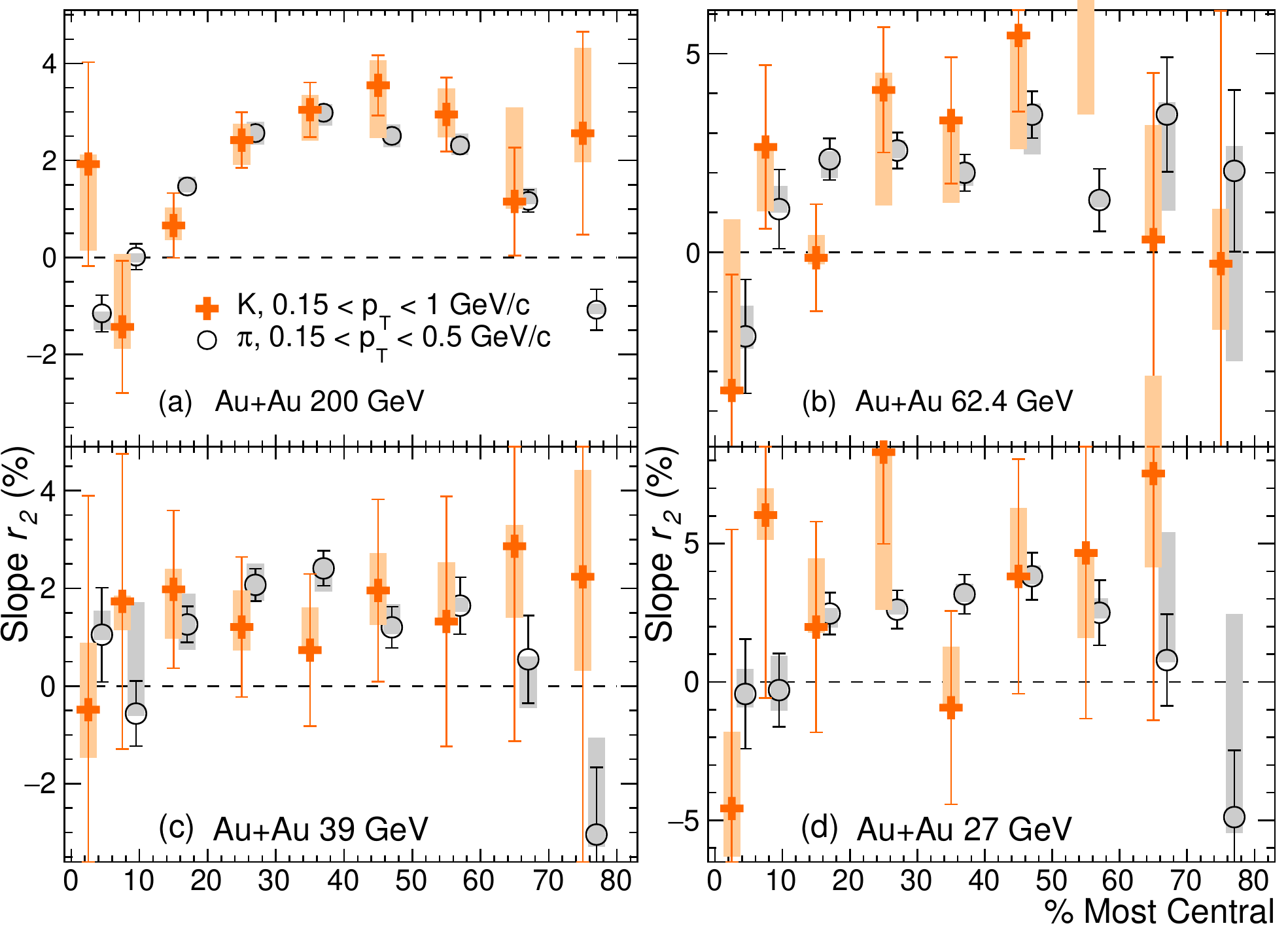}
\caption{Centrality dependence of the \label{label}$r_2$ slopes for kaons and pions in Au+Au collisions at four collision energies.}
\label{fig:slo_bes_k}
\end{figure}


\subsection{Centrality and collision energy dependence of the $r_2$ slope for kaons and (anti)protons}

The prediction of the CMW or LCC effect on $v_2$ for $K^\pm$ and $p$ ($\bar p$) is not as clear as that for $\pi^\pm$. With the same electric quadrupole moment of the QGP, or with the same conditions required by the LCC interpretation, kaons and protons could have a $v_2$ splitting between particles and antiparticles that is weaker than that for pions. One reason is that the differences in the absorption cross sections between $K^+$($p$) and $K^-$($\bar p$) are larger than that of $\pi^\pm$ in the hadronic stage~\cite{Burnier_2011}, which could affect or mask the $\Delta v_2$ from the initial stage. On the other hand, the aforementioned model~\cite{Hatta_2016} with standard viscous hydrodynamics and certain assumptions on the isospin and strangeness chemical potentials ($\mu_{\rm I}$ and $\mu_{\rm S}$) predicts a stronger $v_{2}$ splitting in reverse order for $K^\pm$ (and even stronger for protons) than $\pi^\pm$. In other words, kaons and protons are predicted to have negative $r_2$ slopes with larger magnitudes than pions.. The latter theory is able to successfully reproduce the STAR data of the $v_{2}$ difference between $\pi^-$ and $\pi^+$~\cite{Adamczyk_2013} as well as $r_2$ for pions~\cite{Adamczyk_2015}. Hence the measurements of $r_2$ for $K^{\pm}$ and $p$ ($\bar p$) provide an important test for these physics scenarios.

Figure~\ref{fig:linear_k_p} presents $v_2$ for $K^\pm$ (a), $v_2$ for $p$ ($\bar p$) (b) and the associated $\Delta v_{2}$ (c) as a function of $A_{\rm ch}$ in 30--40\% centrality Au+Au collisions at 200 GeV. We first discuss kaons. The $p_{T}$ range for pions and kaons in this analysis is $0.15 < p_T < 1$ GeV/$c$ as opposed to $0.15 < p_T < 0.5$ GeV/$c$ for pions in the previous STAR publication~\cite{Adamczyk_2015}.  
The wider $p_{T}$ range samples the kaon statistics more efficiently, and meanwhile keeps $\langle p_{T} \rangle$ reasonably flat as a function of $A_{\rm ch}$. Similar to the pion case, the relative variation of the kaon $\langle p_{T} \rangle$ ($\sim$0.1\%) is smaller than that of the kaon $v_{2}$ ($\sim$1\%) by an order of magnitude. Therefore, the $\langle p_{T} \rangle$ effect plays a negligible role in the $r_2$ slope for kaons. 

The prediction by the viscous hydrodynamics model~\cite{Hatta_2016} is contradicted by the measured $r_2$ for kaons, which is positive and close to the pion slope. Note that $\Delta v_2$ at zero $A_{\rm ch}$ is negative for kaons, and positive for pions  with a smaller magnitude. These different $v_2$ orderings qualitatively corroborate the previous observation of the $v_2$ splitting between particles and antiparticles for different species~\cite{Adamczyk_2013}. 

\begin{figure}
\includegraphics[width=20pc]{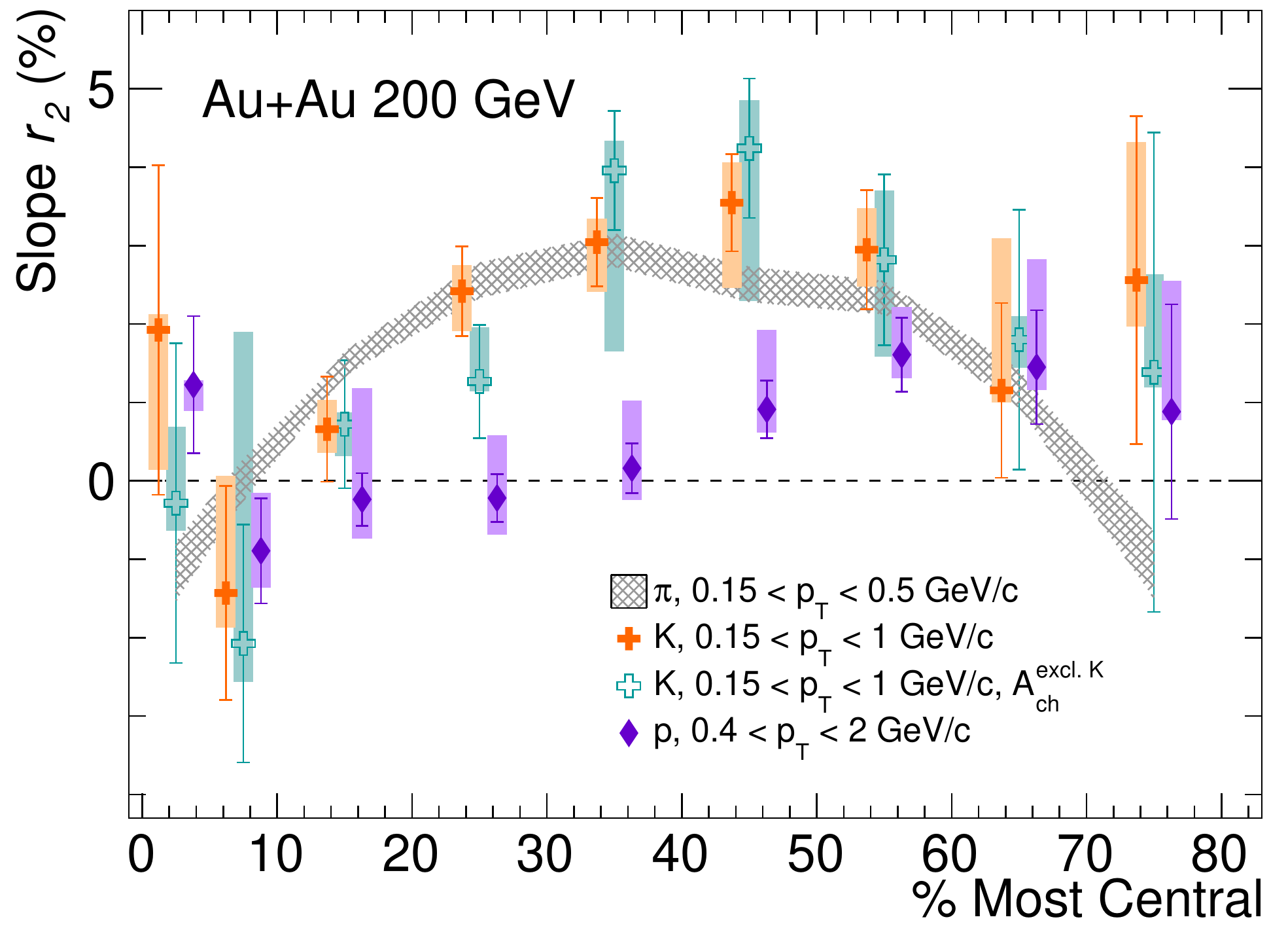}
\caption{Centrality dependence of the $r_2$ for kaons and protons in Au+Au collisions at $\sqrt{s_{\rm NN}}$ = 200 GeV.}
\label{fig:slo_achNoK_p}
\end{figure}

Figure~\ref{fig:slo_bes_k} shows the centrality dependence of the kaon slope in Au+Au collisions at four beam energies: 200, 62.4, 39 and 27 GeV. At $\sqrt{s_{\rm NN}}$ = 200 GeV, the kaon slope displays a rise-and-fall trend, consistent with the pion slope. This consistency holds true for lower energies down to 27 GeV, but the increasing statistical uncertainties do not allow a solid conclusion on the trend. The measurements of the kaon slope do not reveal a significant absorption effect, and suggest that  hydrodynamics with the $\mu_{\rm I}$ and $\mu_{\rm S}$ effects included~\cite{Hatta_2016} cannot be the dominant mechanism for the kaon data. In order to test the trivial ``self-correlation", an additional study is performed by excluding $K^\pm$ from the calculation of $A_{\rm ch}$. The result, as shown in Fig.~\ref{fig:slo_achNoK_p}, is consistent with the default $\pi$ and $K$ results within uncertainties, indicating that the ``self-correlation" effect, if any, is insignificant. 

Compared with kaons, (anti)protons are presumably more affected by the absorption cross section and isospin chemical potential. Moreover, since (anti)protons directly carry the baryonic charge, the chiral vortical effect (CVE)~\cite{Kharzeev_2011} could add to the CME component of CMW for protons, which serves as another potential source of charge separation. The relationship between $\Delta v_{2}$ and $A_{\rm ch}$ for $p$ and $\bar p$ is studied with the same approach as for pions and kaons, except that the $p_{T}$ coverage is $0.4 < p_T < 2$ GeV/$c$, enlarged for the sake of statistics. 
Notably the $v_{2}(A_{\rm ch})$ data for $p$ and $\bar p$ are flatter than those for $\pi^\pm$ and $K^\pm$. The $r_2$ for $p$ and $\bar p$ are thus typically much smaller than those for $\pi^\pm$ and $K^\pm$, as illustrated in Fig.~\ref{fig:linear_k_p} (b). The proton $r_2$ is close to zero for 30--40\% centrality Au+Au collisions at 200 GeV, as shown in Fig.~\ref{fig:linear_k_p} (c).

Figure~\ref{fig:slo_achNoK_p} presents the centrality dependence of the proton $r_2$ slope in Au+Au at $\sqrt{s_{\rm NN}}$ = 200 GeV collisions. 
The proton slopes are close to zero except for the positive values in 40--70\% centrality collisions. The proton data indicate a possible mixed scenario without an obvious dominant mechanism. The contribution of the CMW (CVE) and/or the LCC effect could be reduced by the absorption effect, and/or be counterbalanced by the isospin effect.


\subsection{Centrality dependence of the $r_3$ slope for $\pi^{\pm}$ in Au+Au collisions}

\begin{figure}
\includegraphics[width=20pc]{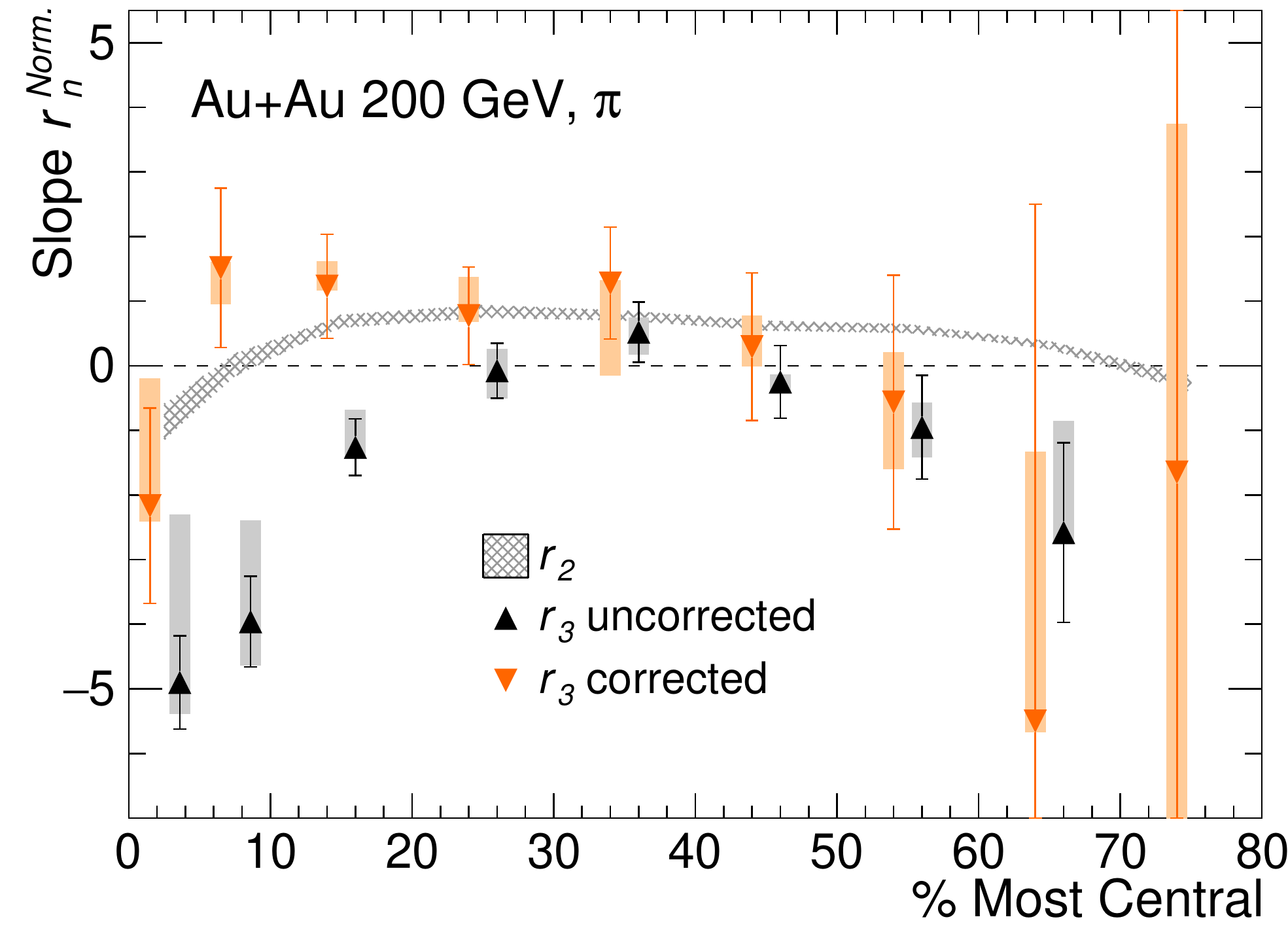}
\caption{The normalized $r_2$ and $r_3$ slopes for pions vs. centrality in Au+Au collisions at $\sqrt{s_{\rm NN}}$ = 200 GeV, with $0.15 < p_{T} < 0.5$ GeV/$c$.}
\label{fig:v2v3}
\end{figure}

In the last section, the kaon data have eliminated the isospin effect from being the dominant contributor to the $r_2$ slopes for pions and kaons, and hence two possible explanations are left: the CMW and the LCC effect. The triangular flow $v_3$ serves as a promising arbitrator, since the LCC effect (and the viscous hydrodynamics calculation with isospin asymmetry) predicts a linear dependence of $\Delta v_{3}$ on $A_{\rm ch}$ for pions, similar to that of $\Delta v_{2}$, whereas the electric quadrupole due to the CMW has no effect on $v_3$. Therefore, the $r_3$ slope, when properly normalized, provides a background estimate for the CMW's contribution to the $r_2$ slope.

Both the $r_2$ and the $r_3$ for pions could be reduced and even go negative owing to a negative correlation between $\Delta v_{2,3}$ and $A_{\rm ch}$, as proposed in Ref.~\cite{Campbell_2013}. Since transported quarks due to baryon stopping suffer more scatterings than produced quarks that are created pairwise in the fireball, the former should bear larger $v_{2,3}$ than the latter. Then assuming pions originate from the coalescence of $u(\bar{u})$ and $d(\bar{d})$ quarks, the $v_{2,3}$ asymmetry between $\pi^+(u\bar{d})$ and $\pi^-(\bar{u}d)$ is determined by the fractions of transported $u/d$ quarks in their constituents. Note that $\bar{u}$ and $\bar{d}$ can only be produced, while $u$ and $d$ can also be transported. To be specific, a positive fluctuation of isospin tends to increase $A_{\rm ch}$ by stopping more protons ($uud$) or fewer neutrons ($udd$), and enriches the $u$ quark population preferentially more than $d$ quarks, which in turn increases  $v_{2,3}$ for $\pi^+$ or decreases that for $\pi^-$. This results in a negative $r_2$ and $r_3$, as confirmed with UrQMD calculations~\cite{Campbell_2013}.

To compare $r_2$ and $r_3$ on the same footing, we normalize $\Delta v_{n}$,
\begin{equation}
\Delta  v_n^{\rm Norm.} = \frac{{\it v_{n}^{-}} - {\it v_{n}^{+}}}{({\it v_{n}^{-}} + {\it v_{n}^{+})}/2},
\label{eq:norm}
\end{equation}
and then extract the normalized $r_2$ and $r_3$ based on $d \Delta v_{n}^{\rm Norm.}/ dA_{\rm ch}$~\cite{Bzdak_2013,PhysRevC.100.064908}.
Figure~\ref{fig:v2v3} compares $r_2^{\rm{Norm.}}$ and $r_3^{\rm{Norm.}}$ for pions as a function of centrality in 200 GeV Au+Au collisions, with the $p_{T}$ range of $0.15 < p_T < 0.5$ GeV/$c$. 
We have used all charged hadrons as RFPs in the analysis, which
could have sizeable nonflow contributions in the $r_n$ slopes,
as pointed out in recent studies~\cite{PhysRevC.101.014913,Xu:2020sln}.
This is because the $d_{n}\{2\}$ in Eq.~(\ref{eq:diff_cumu}) can be decomposed into two terms in accordance with the finite $A_{\rm ch}$~\cite{Xu:2020sln}:
\begin{align}
d^{\pm}_{n}\{2\}
 = & \frac{d_{n}\{2;\pi^{\pm}h^{+}\}+d_{n}\{2;\pi^{\pm}h^{-}\}}{2}  \nonumber \\
& +\frac{d_{n}\{2;\pi^{\pm}h^{+}\}-d_{n}\{2;\pi^{\pm}h^{-}\}}{2}A_{\rm ch}\,. \label{eq:trivial}
\end{align}
The trivial slope in the second term on the r.h.s of Eq.~(\ref{eq:trivial}) is caused by the difference in nonflow correlations between like-sign and  unlike-sign pairs. To eliminate this nonflow effect, one may use positively and negatively charged particles separately as RFPs.
Figure~\ref{fig:v2v3} presents two sets of $r_3^{\rm{Norm.}}$ results: one using all charged hadrons as RFPs (labeled as ``uncorrected''), and the other using positively and negatively charged RFPs separately to extract the $r_3$ slopes which are then combined (labeled as ``corrected'').
It is found that the effect on $r_2$ is relatively small, so we have presented the uncorrected $r_2$ results in this paper. However, significant
systematic differences exist between the corrected and uncorrected $r_3^{\rm{Norm.}}$ values in more central collisions. 
The corrected $r_3^{\rm{Norm.}}$ should be used to compare with $r_2^{\rm{Norm.}}$. 
The $r_2^{\rm{Norm.}}$ and $r_3^{\rm{Norm.}}$ values are expected to be the same if dominated by the LCC effect or the isospin effect. For all centrality intervals under study, the measured $r_3^{\rm{Norm.}}$ is consistent with both zero and $r_2^{\rm{Norm.}}$, within large statistical uncertainties. Therefore, this test with the current precision cannot rule out either the LCC or the CMW scenario.


\subsection{The $r_2$ slopes for $\pi^\pm$ in $p$+Au, $d$+Au and U+U collisions}

\begin{figure}
\includegraphics[width=20pc]{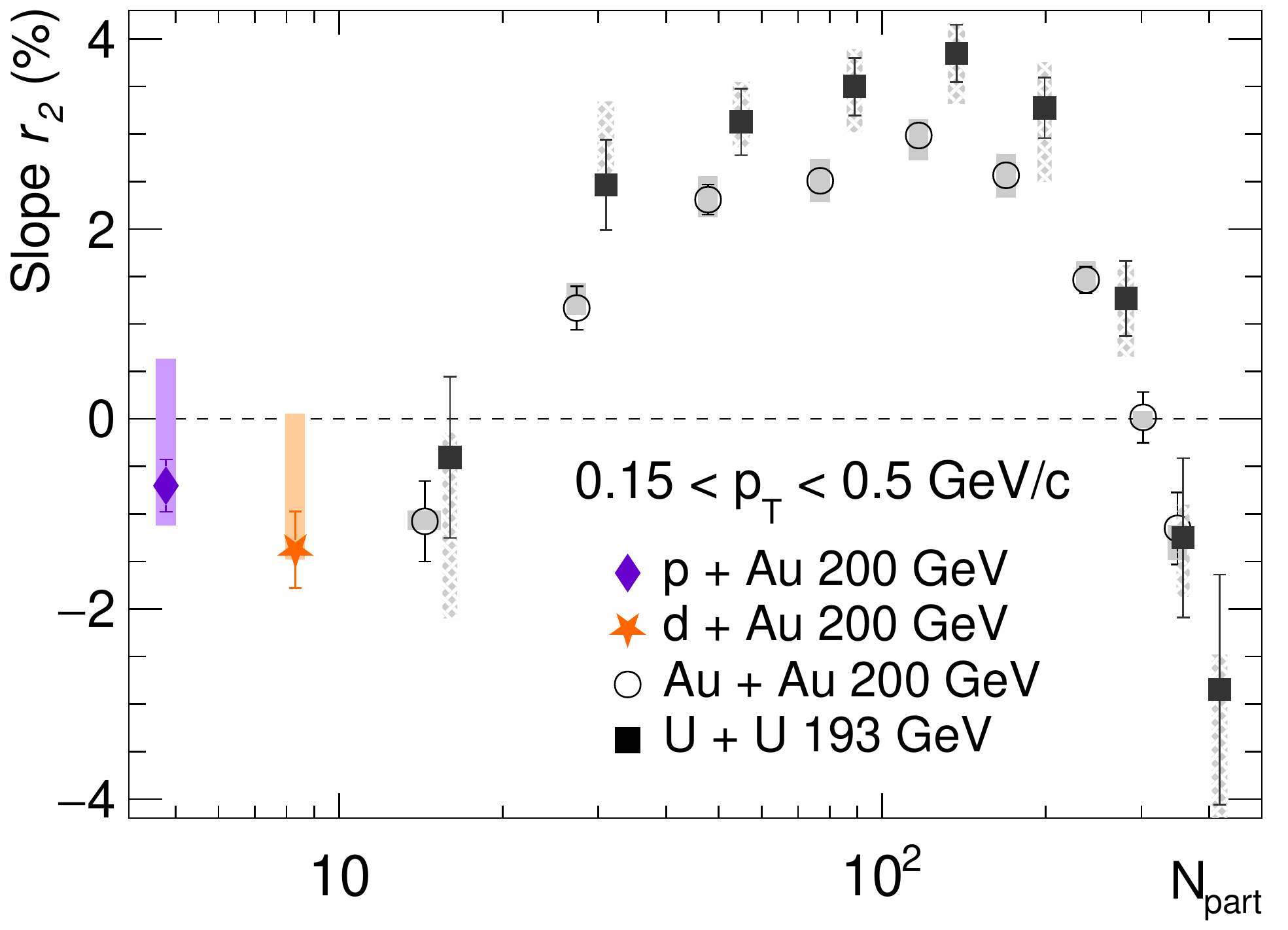}
\caption{The $r_2$ slope for pions vs. $N_{\rm part}$ in $p$+Au, $d$+Au, U+U and Au+Au collisions.}
\label{fig:slo_pA}
\end{figure}

The experimental observation of the CMW relies on the assumption that the direction of the initial intense magnetic field is on average perpendicular to the event plane reconstructed with final-state particles. The angular correlation between the magnetic field and the second-order event plane (reconstructed using the elliptic flow information of particles at midrapidities) is strongest in the intermediate centrality range, and becomes much weaker in central and peripheral collisions~\cite{Bloczynski_2013}. This effect partially explains the rise-and-fall trend observed in the centrality dependence of the $r_2$ slope as shown in previous figures. If the underlying physics is indeed the CMW, then the measured slope parameter is expected to approach zero in very central and very peripheral collisions, where the magnetic field and the second-order event plane both suffer from very strong event-by-event fluctuations, and the correlation between them is almost destroyed. 

In small collision systems such as $p$+Au and $d$+Au, the orientation of the magnetic field is presumably decoupled from the second-order event plane~\cite{Belmont_2017}, which makes such small systems an ideal testing ground for the observation of the disappearance of the $r_2$ slope. 
The CMS measurements~\cite{PhysRevC.100.064908} have observed similarities between the $r_2$ slopes for
$p$+Pb and 
Pb+Pb at 5.02 TeV, supporting the LCC picture and challenging the CMW scenario at LHC energies.
For $p$+Au and $d$+Au collisions at 200 GeV, we present the  pion $r_2$ results (analyzed with the second-order event plane from the TPC) in Fig.~\ref{fig:slo_pA} as a function of $N_{\rm part}$. Data from Au+Au collisions at 200 GeV and U+U collisions at 193 GeV are also shown for comparison. The $r_2$ values in both $p$+Au and $d$+Au are consistent with zero within uncertainties, and corroborate the falling trend previously observed in smaller (peripheral Au+Au) systems. The disappearance of any CMW-like signal in these small systems supports the picture of decoupling between the magnetic field and the second-order event plane~\cite{Belmont_2017}, and demonstrates the smallness of the possible background in the measurement of the CMW signal. The $r_2$ values in U+U collisions are systematically higher than the results in Au+Au collisions in the centrality range where both are prominent. This can be qualitatively explained by the CMW picture, because a uranium nucleus has 13 more protons than a gold nucleus, leading to a stronger magnetic field at the same  $N_{\rm part}$.

\section{Summary} \label{sec:sum}

The previous experimental evidence of the CMW, the $r_2$ slope for pions, has been challenged by interpretations arising from the LCC effect and the isospin effect. In this paper, we present the $r_2$ slopes for low-$p_{T}$ kaons ($0.15 < p_T < 1$ GeV/$c$) in Au+Au collisions at $\sqrt{s_{\rm NN}}$ = 27, 39, 62.4, and 200 GeV. The similarity between pion and kaon slopes suggests that the isospin effect is not the dominant contribution to the pion or kaon slopes. The isospin effect, however, remains a potential contributor to the proton slopes in Au+Au collisions at $\sqrt{s_{\rm NN}} = 200$ GeV. 

The LCC background remains a possible explanation for the positive $r_2$ values for pions and kaons.
The $\langle p_{T} \rangle$ and $\langle |\eta| \rangle$ effect on $A_{\rm ch}$ (and further on $\Delta v_{2}$) has also been discussed in detail,
which seems to be in line with the model study of LCC.
The normalized pion $r_3$ slopes, after correction for nonflow effects, are consistent with both zero and the corresponding normalized $r_2$ slopes within large statistical uncertainties in Au+Au collisions at $\sqrt{s_{\rm NN}} = 200$ GeV.
A much larger data set, {\it e.g.}, from the recent isobar collisions ($^{96}$Ru+$^{96}$Ru and $^{96}$Zr+$^{96}$Zr) at RHIC, is required to draw a firm conclusion on such a test.

The pion $r_2$ slopes have also been reported for $p$+Au and $d$+Au collisions at $\sqrt{s_{\rm NN}}$ = 200 GeV and U+U collisions at $\sqrt{s_{\rm NN}}$ = 193 GeV.  In the small systems, the CMW signals are expected to disappear owing to the orientation decoupling between the magnetic field and the second-order event plane, and the measured slopes are consistent with zero. The difference in the pion $r_2$ slope between Au+Au and U+U is qualitatively consistent with the expectation from the CMW picture. 

Further investigations of the background are needed to draw a firm conclusion on the existence of the CMW in heavy-ion collisions
at RHIC. The large sample of isobar collisions provides just such an opportunity because the magnetic fields are significantly different in the two isobaric systems. 
The second phase of the RHIC Beam Energy Scan program allows for the test of whether the CMW observable  vanishes at low beam energies, \textit{e.g.}, 7.7 GeV, where the partonic interactions are supposed to be dominated by the hadronic ones.
New analysis approaches~\cite{PhysRevC.100.064907,MAGDY2020135986} and new physical mechanisms~\cite{Hong,HONGO2017266,PhysRevC.98.044904,ZHAO2019413,PhysRevC.99.044915} are also proposed to provide further insights into the CMW search.


\section*{Acknowledgements}
We thank the RHIC Operations Group and RCF at BNL, the NERSC Center at LBNL, and the Open Science Grid consortium for providing resources and support.  This work was supported in part by the Office of Nuclear Physics within the U.S. DOE Office of Science, the U.S. National Science Foundation, National Natural Science Foundation of China, Chinese Academy of Science, the Ministry of Science and Technology of China and the Chinese Ministry of Education, the Higher Education Sprout Project by Ministry of Education at NCKU, the National Research Foundation of Korea, Czech Science Foundation and Ministry of Education, Youth and Sports of the Czech Republic, Hungarian National Research, Development and Innovation Office, New National Excellency Programme of the Hungarian Ministry of Human Capacities, Department of Atomic Energy and Department of Science and Technology of the Government of India, the National Science Centre of Poland, the Ministry of Science, Education and Sports of the Republic of Croatia, German Bundesministerium f\"ur Bildung, Wissenschaft, Forschung and Technologie (BMBF), Helmholtz Association, Ministry of Education, Culture, Sports, Science, and Technology (MEXT) and Japan Society for the Promotion of Science (JSPS).

{}


\end{document}